\documentclass{iaus}
\usepackage{graphicx}

\title[The Local Void is Really Empty] 
{The Local Void is Really Empty}

\author[Tully]   
{R. Brent Tully}

\affiliation{Institure for Astronomy, University of Hawaii, Honolulu, HI 96822, USA
\break email: tully@ifa.hawaii.edu}

\pubyear{2007}
\volume{244}  
\pagerange{000--000}
\date{31 July, 2007}
\jname{Dark Galaxies and Lost Baryons}
\editors{J. Davies et al., eds.}
\begin{document}

\maketitle

\begin{abstract}
Are voids in the distribution of galaxies only places with reduced matter
density and low star formation efficiency or are they empty of matter?
There is now compelling evidence of expansion away from the Local Void
at very high velocities. The motion is most reasonably interpreted as
an evacuation of the void, which requires that the void be very large
and very empty.

\keywords{Galaxies: distances and redshifts, intergalactic medium, large-scale
structure of universe}

\end{abstract}

\firstsection 
\section{Introduction}
The dipole pattern in the Cosmic Microwave Background (CMB) gives precise
information about our motion with respect to the mean expansion of the 
universe.  What is the cause of this motion?

Several component influences are well known.  The Sun orbits in our Milky Way 
Galaxy at 220 km/s.  The Milky Way is falling toward Andromeda Galaxy at
135 km/s.  The neighborhood of our Galaxy is retarded from the cosmic
expansion by the Virgo Cluster by 185 km/s, and by large-scale structure
in the direction of the Centaurus constellation by 455 km/s.  But there
must be something else.

The pattern of motions of nearby galaxies reveals the extra component.
We are moving together with all our neighbors within 7~Mpc, the 
{\it Local Sheet}, but there is a
discontinuity in the pattern of velocities as one passes beyond 7~Mpc.
It will be argued that our Local Sheet is repelling from the 
Local Void at 260~km/s! In a $\Lambda$CDM universe with the energy density
parameter $\Omega_{\Lambda}=0.7$, 
a spherical void in an otherwise homogeneous universe evacuates with 
expansion velocities of 16~km/s/Mpc.  The Local Void is inferred to be 
at least 45~Mpc across and really empty.

The following pages will contain a brief outline of the evidence for
these claims.  A more complete discussion has been submitted to the
Astrophysical Journal by Tully, Shaya, Karachentsev, Courtois, Kocevski,
Rizzi, and Peel \cite{tul07}.

\section{Observations}

An accurate distance, $d$, for a galaxy allows us to separate the radial
component of deviant motions, $V_{pec}$, from the cosmic expansion:
$V_{pec} = V_{obs} - {\rm H}_0 d$, where a galaxy has an observed motion
$V_{obs}$ and H$_0$ is the Hubble Constant (taken to be 74~km/s/Mpc).
The present analysis is based on 1797 measured distances for galaxies in
743 groups within 3000~km/s.  In 601 cases, the distance estimates are
based on Cepheid Period-Luminosity \cite{fre01}, Tip of the Red Giant
Branch \cite{sak96}, or Surface Brightness Fluctuation \cite{ton01} methods.
The rest are provided by the correlation between the luminosity of a galaxy
and its line width \cite{tul77}\cite{tul00}\cite{kar02}. 

The large number of accurate TRGB distances to nearby galaxies demonstrates
clearly that the `Local Sheet' of galaxies within 7~Mpc has very low
internal motions in co-moving coordinates \cite{kar03}.  It is shown in 
Figure~1 that our
Milky Way Galaxy has an insignificant random motion with respect to 159
galaxies outside the Local Group but within 7~Mpc.  We are travelling with
the Local Sheet as a unit.

\begin{figure}
\begin{center}
\includegraphics[scale=0.6]{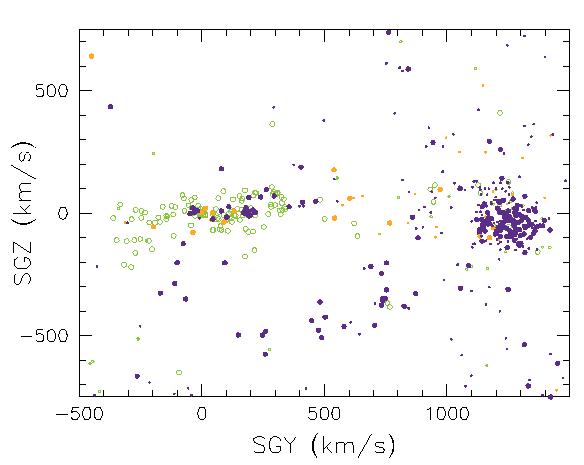}
\caption{Peculiar velocities in the vicinity of the Local Sheet.
Peculiar velocities $<-100$~km/s are coded solid purple (black),
peculiar velocities $>+100$~km/s are coded solid orange (grey), and peculiar
velocities within 100~km/s of zero are coded green (grey) with open symbols.
Our Galaxy lies at SGY=SGZ=0 and the dense clump of galaxies at the right
is the Virgo Cluster.}
\label{ls}
\end{center}
\end{figure} 

However it is seen in Figure~1 and the pan to Figure~2 that there is a
{\it discontinuity} in velocities between the Local Sheet and the next 
nearest structures.  The nature of this discontinuity provides a clue to
the origin of our relative motion.

\begin{figure}
\begin{center}
\includegraphics[scale=0.6]{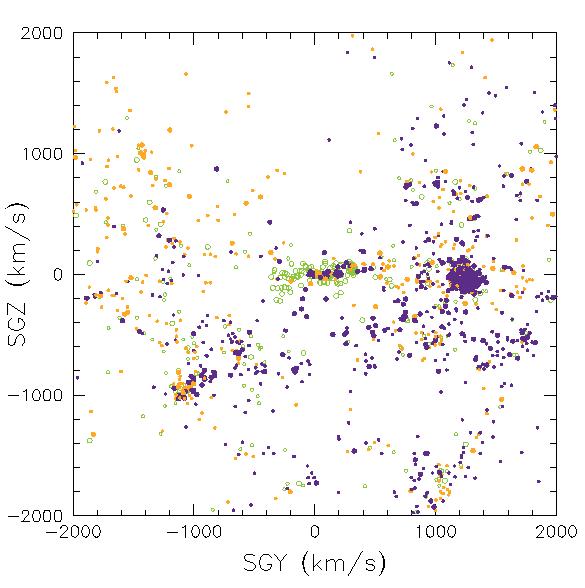}
\caption{Pan to view peculiar velocities within the Local Supercluster.  
Symbol code is the same as in previous figure.}
\label{lsc}
\end{center}
\end{figure} 

The vector of the motion of the Local Sheet with respect to the other
galaxies with distance measures within 3000~km/s but beyond 7~Mpc has an 
amplitude of 323~km/s, toward SGL=80, SGB=--52.  Part of this motion
is suspected to be due to the influence of the Virgo Cluster, 17~Mpc away  
\cite{moh05}. The component of the Local Sheet 
vector toward the Virgo Cluster in the direction 
SGL=102.7, SGB=--2.3 is 185~km/s, consistent with expectations.  
The residual is a 
vector of 259~km/s toward
SGL=11. SGB=--72.  This vector is not directed toward anything of interest,
but it is directed {\it away} from the Local Void.

Following on with this decomposition of the vectors of our motion, we
take note that in the Local Sheet rest frame the CMB dipole vector is
631~km/s toward SGL=139, SGB=--37.  If the motion of the Local Sheet
with respect to galaxies within 3000~km/s is subtracted from the CMB vector,
the residual is 455~km/s toward SGL=162, SGB=--16.  This motion must be 
attributed to influences on scales larger than 3000~km/s.  The vectors
described above are graphically illustrated in Figures 3 and 4.

\begin{figure}
\begin{center}
\includegraphics[scale=0.6]{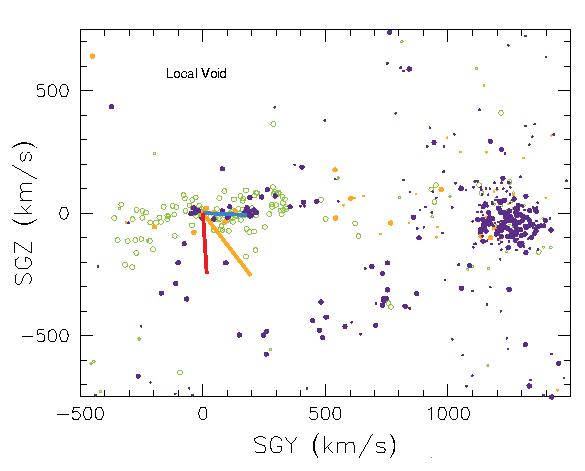}
\caption{Vectors of the motion of the Local Sheet superposed on the data 
of Figure~1. The motion of the Local Sheet with respect to galaxies with
measured distances within 3000~km/s is indicated by the orange (grey) bar 
pointing toward the lower right.  A component of this motion is directed 
toward the Virgo Cluster, indicated by the blue (black) horizontal bar.  
The residual if the component toward Virgo is subtracted from the observed 
motion is the red (black) bar directed downward.
}
\label{vec_lsc}
\end{center}
\end{figure} 

\begin{figure}
\begin{center}
\includegraphics[scale=0.6]{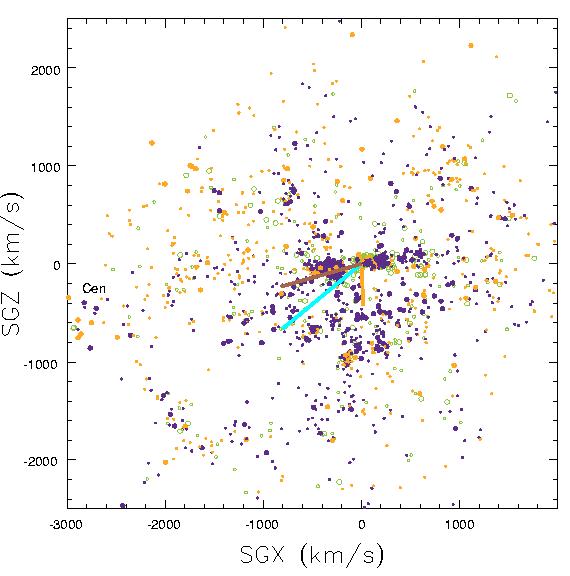}
\caption{Decomposition of the CMB vector, shown in cyan (grey) and directed
from the origin toward the lower left, into the component with respect
to galaxies with velocities $<3000$~km/s, shown in orange (grey) and directed
downward from the origin, and the residual component associated with
structure on scales larger than 3000~km/s, shown in brown (black) and directed
almost horizontal toward the left from the origin. Symbols for individual 
galaxies with distance measures follow the scheme used in previous figures.
The Centaurus Cluster is at SGX=--2800~km/s, SGZ=--500~km/s.
}
\label{vec_lss}
\end{center}
\end{figure} 

\section{The Local Void}

The Local Void was first identified in the Nearby Galaxies Atlas \cite{tul87}.
It is poorly defined because so much of it lies behind the plane of the
Milky Way, but recent surveys such as the HI Parkes All Sky Survey 
\cite{mey04} confirm its general properties.  
There is an underabundance, though not a total lack,
of galaxies in a very large part of the sky at low redshifts.
The empty region begins at the edge of the Local Group, 
with the Local Sheet a bounding surface.  Figure~5 represents an attempt
to illustrate the Local Void.  In detail, there are wispy filaments of galaxies
that lace through the underdense region, causing us to disect it into a
nearby part and more distant `north' and `south' parts.

\begin{figure}
\begin{center}
\includegraphics[scale=0.66]{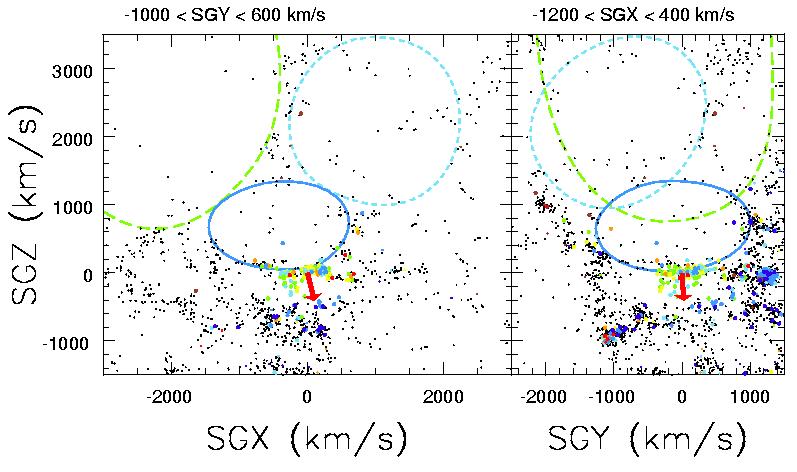}
\caption{The region of the Local Void.  The ellipses outline three apparent
sectors of the Local Void.  The solid dark blue ellipse shows the projection
of the nearest part of the Local Void, bounded at our location by the Local
Sheet.  North and South extensions of the Local Void are identified by the
light blue short--dashed ellipse and the green long--dashed ellipse,
respectively.  These sectors are separated by bridges of wispy filaments.
The red vector indicates the direction and amplitude of our motion away from
the void. 
}
\label{void}
\end{center}
\end{figure} 

Analytic calculations of the expansion of an underdense spherical region
in an otherwise uniform distribution of matter are discussed in the refereed
article.  It is found that an entirely empty region is evacuated at
16~km/s/Mpc.  N-body models provide confirmation \cite{sch07}.  To generate
the observed motion of 260~km/s away from the Local Void, we infer that the 
void at our doorstep must have a diameter of at least 45~Mpc and be very
empty.

It may seem incredible that the CMB dipole motion of the Local Group can 
be decomposed into just three primary contributions: one away from the 
Local Void, a second toward the Virgo Cluster, and a third toward 
large-scale structure in the direction of Centaurus.  Of course, this is 
a simplification.  The reason the three contributions are distinct is because
they arise on distinct scales and are almost orthogonal in direction.

Our attention here is on the nearest component, the motion away from the
Local Void.  There is ongoing discussion about whether places where nothing
is seen are truly empty.  The Local Void generates a `push' of 260~km/s.
Perhaps we are seeing here the best evidence available that voids in the
distribution of galaxies are really empty.

\begin{acknowledgments}
My collaborators in this research are Ed Shaya, Igor Karachentsev,
H\'el\`ene Courtois, Dale Kocevski, Luca Rizzi, and Alan Peel.
Support has been provided by grants from Space Telescope Science Institute
and the US National Science Foundation.
The motion of our Galaxy with respect to the structure in the distribution
of galaxies is illustrated in videos at
http://www.ifa.hawaii.edu/~tully/
\end{acknowledgments}

\end{document}